\documentclass[aps, prl, reprint, preprintnumbers, floatfix, superscriptaddress, longbibliography]{revtex4-2}
\pdfoutput=1

\usepackage{amsmath, amssymb}

\usepackage{graphicx}
\usepackage[dvipsnames]{xcolor}

\usepackage{booktabs}

\usepackage{siunitx}
\usepackage{mathtools} \usepackage{derivative} 

\usepackage{xspace}

\usepackage{hyperref}
\hypersetup{
breaklinks=true,
colorlinks=true,
linkcolor=blue,
filecolor=magenta,      
urlcolor=Mulberry,
citecolor=magenta,
pdftitle={Space-time Floquet operator}
}

\AtBeginDocument{
  \heavyrulewidth=.08em
  \lightrulewidth=.06em
  \cmidrulewidth=.03em
  \belowrulesep=.65ex
  \belowbottomsep=0pt
  \aboverulesep=.4ex
  \abovetopsep=0pt
  \cmidrulesep=\doublerulesep
  \cmidrulekern=.5em
  \defaultaddspace=.5em
  }

\newcommand{\papertitle}{Space-time Floquet operator: Nonreciprocity and fractional topology of space-time crystals}
\newcommand{\acktext}{This work was supported by the National Science Foundation under Grant No. DMR–2145766.}

\DeclarePairedDelimiter\parens{\lparen}{\rparen}

\DeclarePairedDelimiter\bracks{\lbrack}{\rbrack}

\newcommand{\iunit}{\mathrm{i}\mkern1mu} 

\newcommand{\eqnref}[1]{Eq.~\eqref{#1}}
\newcommand{\Equationref}[1]{Equation~\eqref{#1}}

\newcommand{\subfiglabel}[1]{{(#1)}}
\newcommand{\figref}[1]{Fig.~\ref{#1}}

\newcommand{\subfigref}[2]{\figref{#1}\subfiglabel{#2}}

\newcommand{\vb}[1]{\mathbf{#1}}

\newcommand{\ket}[1]{{| #1 \rangle}}
\newcommand{\bra}[1]{{\langle #1 |}}

\DeclareMathOperator{\Tr}{Tr}

\newcommand\ham{\mathcal{H}}

\newcommand\heff{\mathcal{H}^\textrm{eff}}

\newcommand{\im}{\beta}
\newcommand{\ip}{\alpha}
\newcommand{\rim}{r_\beta}
\newcommand{\rip}{r_\alpha}

\newcommand{\gvec}[1]{\vb g_{#1}}
\newcommand{\gvecF}[1]{\vb g_{#1}^{\textrm{F}}}

\newcommand{\siRefer}{\cite{SupplementalMaterial}} 
\makeatletter\begin{document}

\preprint{\textsf{Published in \textit{Phys. Rev. Research}
  \textbf{8}, L042009 (2026)}}
\preprint{\textsf{\href{https://doi.org/10.1103/xcls-6nzd}
  {doi:10.1103/xcls-6nzd}}}

\title{\papertitle}
\author{Abhijeet Melkani}
\email{abhijeet.melkani@ens-lyon.fr}
\affiliation{Institute for Fundamental Science, Materials Science Institute, and Department of Physics,
University of Oregon, Eugene, OR 97403}
\affiliation{École Normale Supérieure de Lyon, CNRS, Université Claude Bernard Lyon 1, 
Laboratoire de Physique, F-69342 Lyon, France}

\author{Jayson Paulose}
\email{jpaulose@uoregon.edu}
\affiliation{Institute for Fundamental Science, Materials Science Institute, and Department of Physics,
University of Oregon, Eugene, OR 97403}

\begin{abstract}

We introduce a \emph{space-time Floquet operator} for crystals whose spatial and temporal translations are intertwined.
Unlike the conventional Floquet operator, which describes evolution over a full time period, the space-time Floquet operator evolves the system over a fraction of the period paired with a spatial shift that generates the primitive space-time unit cell.
Its eigenmode spectrum defines a space-time band structure that unfolds conventional Floquet bands to reveal the consequences of the intertwined crystal symmetry in reciprocal frequency-wavevector space.
We relate the topology of these space-time bands to quantized transport phenomena, such as Bloch oscillations and adiabatic charge transport, and uncover an intrinsically fractional charge pump enabled by the primitive space-time cycle.
Numerical simulations quantitatively support these predictions with small deviations controlled by the speed of the modulation.
We also demonstrate how nonreciprocal parametric resonances are naturally anticipated and accurately predicted by our framework.
Our approach applies broadly to both classical and quantum systems with space-time symmetry, including non-Hermitian crystals. \end{abstract}

\maketitle

\paragraph{Introduction.}

Periodic drive provides a powerful tool to design unconventional quantum~\cite{Oka2019,Rudner2020} and classical~\cite{Yin2022} non-equilibrium phases, often with no equilibrium counterpart.
In many cases, such driven systems exhibit a discrete symmetry under a combined translation in space and time that does not simply decompose into independent periodicities in the spatial and temporal dimensions.
Such structures, called \emph{space-time crystals}~\cite{xu2018spaceTimeGroup}, are generically created when a static crystal has its local properties modulated by a traveling wave.
Classically, this mixing of space and time underpins reciprocity-breaking in linear response~\cite{nassar2020nonreciprocity} and enables the transfer of energy from the driving field to electromagnetic or mechanical waves via traveling-wave parametric amplification~\cite{Cullen1958, Tien1958, Simon1960, Galiffi2019, kruss2022oneway}.
In quantum condensed matter, space-time crystals are predicted to harbor novel topologically protected states~\cite{Morimoto2017, peng2022spaceTimeCrystal} and nontrivial oscillatory dynamics under external electric fields~\cite{gao2021spaceTimeTunneling}.
However, these rich symmetries pose theoretical and computational challenges for current analysis methods.

For an ordinary lattice subject to a periodic drive, the primitive spatial and temporal translations are independent, and Floquet analysis~\cite{grifoni1998driven,Bukov2015} can be combined with the Bloch framework for spatially periodic systems.
In space-time crystals, by contrast, an irreducible primitive symmetry advances space and time together, while the separate translations by a larger spatial supercell and full modulation period are coarser symmetries that derive from the mixed one.
If a space-time crystal is treated only as a conventional lattice plus periodic drive, the resulting Floquet band structure~\cite{Kolovsky1994,kitagawa2010periodic,gomez-leon_floquet-bloch_2013} provides a route to characterizing the linear response and topological properties of the system~\cite{Rudner2020a}.
However, this Floquet-Bloch approach fails to preserve the primitive space-time symmetry because it involves integrating the dynamics over the full modulation period, missing the finer time-scale set by the intertwined periodicity~\cite{melkani2024resonance}.
The resulting band structures exhibit spurious eigenvalue degeneracies (band crossings) that do not correspond to  physical resonances~\cite{kruss2022oneway, gao2021spaceTimeTunneling, melkani2024resonance} and provide little insight into reciprocity-breaking behavior~\cite{kruss2022oneway, melkani2024resonance}.
Frequency-domain approaches~\cite{Simon1960,xu2018spaceTimeGroup,gao2021spaceTimeTunneling} resolve these issues by using an oblique reciprocal lattice to expand the eigenmodes of the system while explicitly satisfying the space-time symmetry, but actual computations require truncation of an infinite-dimensional coupling matrix in Fourier space~\cite{gomez-leon_floquet-bloch_2013,Simon1960,Rudner2020a}.
The number of harmonics that must be kept for an accurate evaluation of the spectrum can be very large~\cite{Bukov2015} or even infinite~\cite{Hessel1961,Cassedy1963,Cassedy1967} in strongly driven or strongly interacting systems, limiting the utility of the frequency-domain approach.

Here, we resolve these issues by defining a \emph{space-time Floquet operator}.
Unlike the Floquet operator, which captures the evolution of a state over a complete modulation period $T$, the operator we define generates time-evolution over an interval $T/\beta$ determined by the spatiotemporal symmetries of the crystal~\cite{melkani2024resonance}.
We define this operator by building on our finite-system framework introduced in Ref.~\cite{melkani2024resonance} to construct its Bloch-resolved form for extended lattices and general spatiotemporal symmetry.
Our results supersede both Floquet-Bloch and frequency-domain approaches by providing a framework based on direct time-evolution that avoids \emph{ad hoc} truncation schemes while respecting the oblique reciprocal lattice structure.

\begin{figure}[t]
\centering
\includegraphics{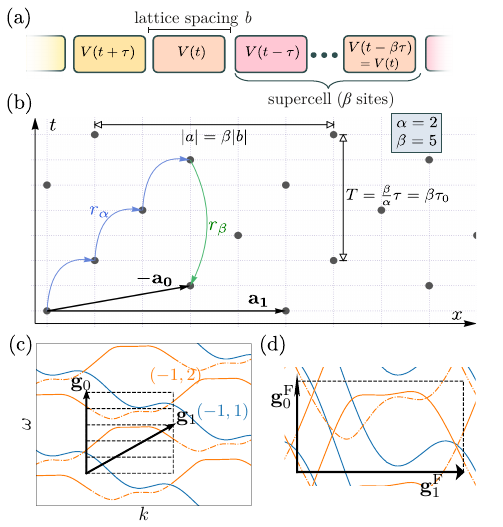}
\caption[]{\label{fig:fig1} 
  \subfiglabel{a} Schematic of a $1$D tight-binding system with space-time periodic onsite potential $V(t)$.
  \subfiglabel{b} Space-time periodic lattice with $\im=5$ and $\ip=2$ showing the primitive vectors ($\vb a_0$, $\vb a_1$) and the geometric relationships between the quantities $\rim$ and $\rip$ (here, $-1$ and $3$ respectively), the time period $T$ and the smallest time-step $\tau_0$.
\subfiglabel{c} Space-time band structure on the reciprocal lattice generated by primitive vectors ($\vb g_0$, $\vb g_1$). Two bands are shown with their associated windings; the orange band combines two Bloch states with distinct indices (solid and dashed).
  \subfiglabel{d} The Floquet band structure collapses $\beta = 5$ plaquettes [dashed boxes from (c)] onto a smaller Floquet-Bloch Brillouin zone spanned by ($\gvecF{0},\gvecF{1}$).
}
\end{figure}

\paragraph{Space-time Floquet operator.}\label{sec:theory}
While our framework holds for systems in any spatial dimension as long as they satisfy periodic boundary conditions (PBC),
we frame our discussion using a $1$D system with a Hamiltonian of the form $\mathcal{H}(t) = \frac{p^2}{2m} + V(x,t)$ such as in \subfigref{fig:fig1}{a}.
The system has space-time symmetry~\cite{xu2018spaceTimeGroup, gao2021spaceTimeTunneling, peng2022spaceTimeCrystal} if it satisfies $V(x + b, t+ \tau) = V(x,t)$.
That is, a traveling wave with velocity $v = \frac{b}{\tau}$ modulates the system.
The time-step $\tau$ is taken to be positive while the sign of $b$ is set by the direction of the traveling wave
~\footnote{
The sign of $b$ is chosen such that the associated $\tau$ is smaller. 
For example, the equation $V(x + b, t+ \tau) = V(x,t)$ implies that $V(x, t) = V(x - b, t - \tau) = V(x - b, t +\frac{\im - \ip}{\ip}\tau)$. 
We require that $\frac{\im - \ip}{\ip} > 1$. 
If $\frac{\im-\ip}{\ip} < 1$, we reverse the direction of $b$. 
If $2\ip = \im$ (which only occurs when $\im=2$ and $\ip=1$), the choice of sign of $b$ does not matter as the left-right symmetry is unbroken and there is no traveling wave modulation.}.
The displacement $|b|$ is the shortest length for which such an equation may be written.

We denote the system size by $L:= n \im |b|$, where $n$ is the number of supercells in the periodic system and $|a|:= \im |b|$ the size of each supercell ($\im$ is a positive integer).
If $n>1$ we have a lattice and when $n=1$ we have a space-time symmetric ring~\cite{melkani2024resonance} without translational symmetry.
Since $V(x,t) = V(x+ \im b, t) = V(x, t - \im \tau )$, the potential $V(x, t)$ (and by consequence, the Hamiltonian) is required to be time-periodic with some time-period $T$ whose general form is $T= \frac{\im}{\ip}\tau$, with $\ip$ a positive integer coprime with $\im$~\footnote{
The integer $\ip$ can always be chosen to be smaller than $\im$ which we will assume is the case.
If $\ip$ is larger than $\im$ replace $\ip$ with $\ip \operatorname{mod} \im$.
}.

While the space-time Bravais lattice of the crystal is spanned by any pair of independent lattice vectors, we make a special choice of primitive vectors involving the \emph{shortest time-step} needed to capture the long-time evolution of the system, $\tau_0 := \frac{T}{\im} = \frac{\tau}{\ip}$ [graphically represented in \subfigref{fig:fig1}{b}].
We use B\'{e}zout's identity to find the integers $\rim$ and $\rip$ satisfying $\rim \im + \rip \ip=1$~\footnote{
All such solutions are of the form $(r_\beta, r_\alpha) \in (\rim^0, \rip^0) + \mathbb{Z}(\ip, -\im)$ with $(\rim^0, \rip^0)$ a particular solution.
We choose the one such that $ 0 < \rip < \im$. 
} 
such that
\begin{equation}
V(x,t) = V(x + \rip b, t + \rip \tau + \rim T) = V(x + \rip b, t + \tau_0).\nonumber
\end{equation}
The requisite primitive vectors are then $\vb a_0 = -(\rip b, \tau_0)$ and $\vb a_1 = (\beta b, 0)$ [\subfigref{fig:fig1}{b}]. 
These quantities, together with the translation operator $S(r) := e^{\iunit rp}$ (we set $\hbar = 1$ throughout), allow us to express the periodicity conditions on the Hamiltonian as
\begin{subequations}
\begin{align}
\ham(t) &= S(a)\ham(t)S^{-1}(a), \label{eq:translationalSymmetry}\\
\ham(t) &= \ham(t+T), \\
\ham(t) &= S(\rip b)\ham(t+\tau_0)S^{-1}(\rip b), 
\end{align}
\end{subequations}
corresponding to translational symmetry over the supercell, Floquet symmetry, and space-time symmetry respectively.
The first two conditions are the symmetries of an ordinary periodically driven lattice, and constrain the regular Floquet band structure.
The third condition is the additional space-time constraint: evolution by the fractional time step $\tau_0$ returns the Hamiltonian to itself only after a spatial translation, so it cannot be captured by treating the drive and lattice periodicity as independent.
The time-propagation operator generated by the Hamiltonian evolution, $U(t):= \mathcal{T} e^{-\iunit \int_0^t \odif{t'} \ham (t')}$, 
inherits these symmetries and satisfies
\begin{subequations}
\begin{align} 
U(t) &= S(a)U(t)S^{-1}(a), \\
  U(t+T) &= U(t)U(T), \\
  U(t+ \tau_0) &= S^{-1}(\rip b) U(t) S(\rip b) U(\tau_0). \label{eq:spaceTimeFloquet}
\end{align}
\end{subequations}

\Equationref{eq:spaceTimeFloquet}, which we prove in \siRefer, shows explicitly that evolution over the shortest time-step is sufficient to reproduce the long-time behavior of the system.
Specifically, it implies that the evolution over $m$ integral time steps is
\begin{equation}
U(t + m \tau_0) = S^{-m}(\rip b)U(t) [S(\rip b) U(\tau_0)]^m\label{eq:strongerThanFloquet}.
\end{equation}
This form is reminiscent of the Floquet form for evolution in multiples of the time period of the potential, $U(t+ mT) = U(t) [U(T)]^m$ where $U(T)$ is the Floquet operator. 

To proceed, we transform to a Bloch basis to account for the spatial periodicity.
We change basis to the eigenstates of $S(a)$ which simultaneously block-diagonalizes the operators $\ham(t)$, $U(t)$, and $S(\rip b)$ into $N$-dimensional blocks where $N$ is the number of system degrees of freedom in a supercell of size $a$.
Each block of $S(a)$ is given by $S_k(a) = e^{ika}\mathbb{I}_N$ where $k \in (-\frac{\pi}{a}, \frac{\pi}{a} ]$ is the wavevector in reciprocal space.
Using this basis, \eqnref{eq:spaceTimeFloquet} is represented as (see \siRefer for details)
\begin{align}
U_k(T) = U_k\parens*{\im \tau_0} = e^{-\iunit \rip k a } \bracks*{S_k\parens*{\rip b} U_k\parens*{\tau_0} }^\im,
\end{align}
where $U_k(T)$ is the Floquet operator in wavevector space (whose spectrum generates the conventional Floquet-Bloch band structure) and we used $S_k^{-\beta}(\rip b) = S_k^{-1}(\rip a) = e^{-\iunit \rip k a} \mathbb{I}_N$.
Taking the principal $\im$th root of this relation, we define
\begin{equation}\label{eq:defineSTcont}
	X_k := e^{-\iunit \rip k b} S_k(\rip b) U_k(\tau_0)
\end{equation}
as the $k\textrm{th}$ block of the \emph{space-time Floquet operator} which factorizes the Bloch-transformed Floquet operator as
\begin{equation}\label{eq:Floquetfactorize}
U_k(T) = X_k^\im.
\end{equation}

Following the approach used to generate Floquet band structures, we define the $N$-dimensional effective Hamiltonian $\heff_k$ of the system by $e^{-\iunit \heff_k \tau_0}:=X_k$ and denote its $j\textrm{th}$ eigenvalue by $\omega_j(k)$.
Henceforth, we will refer to the quantum number $j$ as the state index of the eigenstate.
The $N$ functions $\omega_j(k)$ define the \emph{space-time band structure} for space-time crystals. Each space-time frequency eigenvalue $\omega_j(k)$ is defined modulo $\frac{2\pi}{\tau_0} = \frac{2\pi \im}{T}$.
The corresponding eigenvectors $\Psi_{k,j}(x)$ are Bloch states satisfying $\Psi_{k,j}(x + a) = e^{\iunit k a}\Psi_{k,j}(x)$.
A state initialized in an eigenvector $\Psi_{k,j}(x)$ at $t=0$ evolves over an integer multiple of the shortest time-step as 
\begin{equation} \label{eq:stroboscopic}
U_k(m \tau_0) \Psi_{k,j}(x) = e^{\iunit m(r_\alpha k b - \omega_j \tau_0)} S_k^{-m}(\rip b)  \Psi_{k,j}(x);
\end{equation}
\emph{i.e.}, the stroboscopic dynamics of the eigenstate at intervals of duration $\tau_0$ are generated simply by appropriate translations in space.
This feature translates to the stroboscopic dynamics of arbitrary initial states when the  $\Psi_{k,j}(x)$ form a complete basis for the respective Bloch subspace, such as when $\ham_k(t)$ is Hermitian.

The space-time band structure inherits the intertwined space-time translational symmetries in the reciprocal wavevector-frequency space.
The space-time lattice depicted in \subfigref{fig:fig1}{b} is associated with a reciprocal lattice with primitive vectors $\vb g_i$ defined via $\vb g_i \cdot \vb a_j = 2\pi \delta_{ij}$~\cite{xu2018spaceTimeGroup, gao2021spaceTimeTunneling}.
For our choice of direct primitive vectors, these evaluate to $\gvec{0} = \left(0, \frac{2\pi}{\tau_0} \right)$ and $\gvec{1} = \left(\frac{2\pi}{\beta b},\frac{2\pi r_\alpha}{\beta \tau_0} \right)$,
signifying an oblique reciprocal lattice that mixes frequency and wavevector directions.
Space-time symmetry ensures that the space-time band structure $\omega_j(k)$ generates a periodic tiling of the frequency-momentum space with these primitive vectors, as shown schematically in \subfigref{fig:fig1}{c} for a system with two unique non-crossing bands.

\Equationref{eq:Floquetfactorize} shows that the space-time bands (the spectrum of $X_k$) are closely related to Floquet bands (the spectrum of $U_k(T)$) but the latter do not capture the finer space-time symmetric structure.
Conventional Floquet-Bloch analysis treats the temporal and spatial periodicities of the space-time crystal, with periods $T = \beta \tau_0$ and $a = \beta b$ respectively, as independent~\cite{kitagawa2010periodic,gomez-leon_floquet-bloch_2013,Holthaus2015,Rudner2020}.
This construction is appropriate for a generic periodically driven lattice, but for a space-time crystal it discards the primitive mixed translation. The primitive vectors $\vb a_0^{\textrm{F}} = -(0, \beta \tau_0)$ and $\vb a_1^{\textrm{F}} = (\beta b, 0)$ generate a rectangular reciprocal lattice spanned by $\gvecF{0} = \left(0, \frac{2\pi}{\beta \tau_0} \right)$ and $\gvecF{1} = \left(\frac{2\pi}{\beta b}, 0 \right)$.
The Floquet bands $\omega^{\textrm F}_j(k)$ are the eigenvalues of $\frac{i}{T}\log{U_k(T)}$; by \eqnref{eq:Floquetfactorize}, they ``fold'' the space-time bands onto a narrower frequency range via $\omega^{\textrm F}_j(k) = \omega_j(k) \mod \frac{2\pi}{T}$ (\subfigref{fig:fig1}{d}), giving rise to spurious band crossings that do not reflect actual eigenvalue couplings as they are not present in the space-time band structure.
The corresponding eigenvectors $\Psi_{k,j}^\textrm{F}(x)$ are also Bloch states with period $a$, but do not necessarily satisfy the finer structure of \eqnref{eq:stroboscopic}.

\begin{figure*}[tb]
  \centering
  \includegraphics{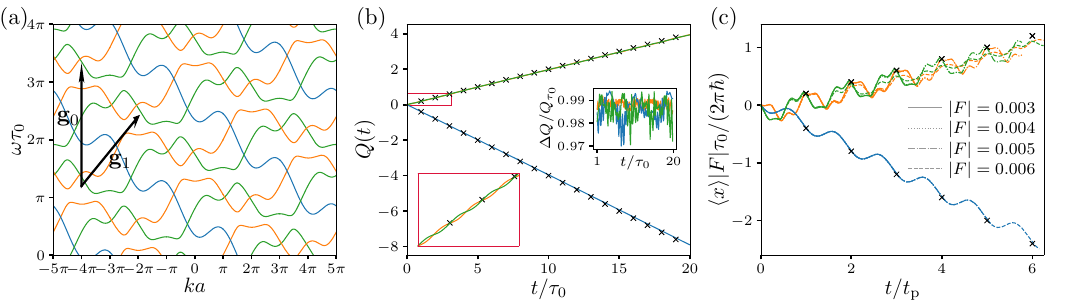}
  \caption{
\subfiglabel{a} Space-time band structure of  Hamiltonian in \eqnref{eq:tbham} with $\im = 5$, $\ip =2$, $V_0 = 1.5$, $J = 1$, $T = 29$.
    Blue, orange, and green bands have windings $(w\parens*{\gvec{0}},w\parens*{\gvec{1}}) = (-1,1)$, $(-1,2)$ and $(-1,2)$ respectively.
Band assignments and quantized transport predictions neglect the minuscule gaps at the crossing points, the largest of which is $\Delta\omega \approx 0.0119/\tau_0$, and treat these points as true crossings (see \siRefer).
\subfiglabel{b} Net pumped charge $Q(t)$ for filled bands, obtained by numerically integrating Schr\"odinger's equation on a chain with $n=400$. Colors indicate the bands from (a).
    Crosses show the quantized expectation $Q(n\tau_0) = nQ_{\tau_0}$ at integer multiples of $\tau_0$.
    Left inset: Zoom of boxed region to show detail.
    Right inset: Pumped charge $\Delta Q(t) = Q(t) - Q(t-\tau_0)$ normalized by the winding prediction. 
    \subfiglabel{c} Shift in center of mass during Floquet-Bloch oscillations for wavepackets initialized using eigenstates of the distinct bands from (a) (colors) at three different values of the external field (dashing) in a chain with $n=200$.
Crosses show the accumulation of the predicted shift over one oscillation, \eqnref{eq:blochosc}.
  }
  \label{fig:sims}
\end{figure*}

\paragraph{Topological phenomena.}\label{sec:winding}
For any point $(k,\omega)$ on a band in the space-time band structure, we can move along the band to its first lattice copy, $(k, \omega) + w(\gvec{0}) \gvec{0} + w(\gvec{1}) \gvec{1}$.
The integers $w(\gvec{0})$ and $w(\gvec{1})$ define the winding numbers of the band~\footnote{Since we follow the band along increasing values of $k$, $w(\gvec{1})$ is a positive integer}.
A key advance of our operator construction is a unified formulation in which the two windings $w(\gvec{0})$ and $w(\gvec{1})$ determine both quantized transport and field-driven wavepacket dynamics.
This formulation builds on earlier studies of oblique space-time bands~\cite{xu2018spaceTimeGroup, gao2021spaceTimeTunneling} and generalizes established winding-based characterizations of Floquet bands~\cite{kitagawa2010periodic, gomez-leon_floquet-bloch_2013, Rudner2020} to the space-time symmetric context.

In Hermitian systems, Floquet band topology relates to quantized adiabatic charge transport or ``Thouless pumping'' under time-modulated potentials~\cite{Thouless1983a,Citro2023}.
In a conventional Floquet analysis, the natural closed cycle is the full temporal period $T$, and the current in a filled Floquet band integrated over this period is quantized to an integer given by the winding number of the band in the periodic frequency direction~\cite{kitagawa2010periodic}.
Space-time symmetry supplies an alternate closed cycle---evolution over the shortest time-step $\tau_0=T/\beta$ followed by the spatial translation appearing in $X_k$---which correctly captures the mixed primitive cell.
Upon integrating the spatial average of the instantaneous current~\cite{kitagawa2010periodic} of a filled band over this time-step, we obtain (see \siRefer)
\begin{align}
Q_{\tau_0} := \int_0^{\tau_0} \odif t \, J(t) &= \int_{\mathrm{BZ}} \! \frac{\odif k}{2\pi}\iunit \Tr \bracks*{P_k X^{-1} _k \pdv{X_k}!{k} }\nonumber\\
	&= w\parens*{\gvec{0}} + \frac{\rip}{\im} w\parens*{\gvec{1}},\label{eq:quantizedCurrent}
\end{align}
where the trace is over the full Bloch subspace, with the projector $P_k$ selecting the occupied eigenstates.
Thus the charge pumped across the system over a time interval $\tau_0= T/\beta$ is topologically quantized to a rational fraction, rather than an integer.

The fractional quantization of pumped charge revealed by the space-time Floquet operator is not merely the result of slicing an ordinary Floquet pump into $\beta$ time windows: after each interval $\tau_0$, the state returns to an equivalent point in the space-time crystal only after the accompanying spatial translation.
Fractional charge pumping was previously proposed for systems with special symmetries by partitioning the integral of the Berry curvature of instantaneous eigenstates over periodic momentum and time into $\beta$ equal parts~\cite{marra2015fractional}.
In contrast, \eqnref{eq:quantizedCurrent} derives the fractional response directly from the two windings of the space-time bands defined by $X_k$ on the oblique reciprocal lattice.
It recovers Thouless's integer result over a full period~\cite{Thouless1983a} (see \siRefer), while identifying the topological contribution of a single primitive space-time cycle that will also control the Floquet-Bloch dynamics below.

To test the prediction of fractional charge quantization, we simulated wavepacket dynamics in a one-dimensional tight-binding chain with a discrete space-time symmetric Hamiltonian
\begin{multline}
  \label{eq:tbham}
  \mathbf{H} = \sum_j \biggl[-J|j\rangle \langle j+1 | -J  |j\rangle \langle j-1 | \\
    + V_0 \cos \left(\frac{2\pi\alpha}{\beta} j -\frac{2\pi}{T}t \right)  |j\rangle \langle j | \biggr].  
\end{multline}
In the resulting space-time band structure, shown in \subfigref{fig:sims}{a}, the five degrees of freedom per Bloch subspace appear to organize into three contiguous bands that cross each other as they wind around the Brillouin zone.
Upon closer inspection (Supplementary Figure S1), the crossings are revealed to be weakly avoided crossings with gaps $\Delta \omega$ of order $0.01/\tau_0$ 
that disrupt the windings of the space-time bands.
However, these gaps are expected to fall exponentially with the modulation period $T$~\cite{Lindner2017}, producing small nonadiabatic corrections to the idealized quantized pumping~\cite{Shih1994,Privitera2018}.
Accordingly, we ignore the miniscule gaps at avoided crossings to organize the bands into effective diabatic branches with nontrivial windings (\subfigref{fig:sims}{a}).
For this diabatic band assignment, \eqnref{eq:quantizedCurrent} is no longer exact, but rather is accurate only up to nonadiabatic corrections of order $\tau_0 \Delta \omega \sim 0.01$ (see \siRefer).

We computed the net pumped charge associated with wavepackets assembled from all Bloch states of each diabatic band (see \siRefer for simulation details).
The pumped charge $Q(t)$ at intervals $t=n\tau_0$ closely tracks integer multiples of the fractional quantized charge predicted by \eqnref{eq:quantizedCurrent}, which has values $Q_{\tau_0} = 1/5$ for the green and orange bands and $Q_{\tau_0} = -2/5$ for the blue band (\subfigref{fig:sims}{b}).
The deviation from the exact quantization over each cycle is 3\% or less (inset), consistent with the expected correction due to the weakly avoided crossings.
These simulations show that the fractional quantization predicted by \eqnref{eq:quantizedCurrent}, exact for isolated space-time bands, can be observed with small quantifiable corrections for realistic systems with weakly avoided crossings.

The periodicity of the space-time band structure in momentum and frequency directions causes an electron to move back and forth in response to a unidirectional electric field, in a phenomenon termed Floquet-Bloch oscillations~\cite{gao2021spaceTimeTunneling}, which we also relate to winding topology.  In the presence of a constant external force $F$, weak enough that the resulting dynamics are slow compared to the time scales in the band structure, the adiabatic evolution of the quasimomentum of each state is well-described by $k(t) = k(0) + \frac{F}{\hbar}t$~\cite{bendahan1996oscillations,gao2021spaceTimeTunneling}. 
As the momentum-frequency pair traverses the band and returns to successive lattice images of itself, the group velocity of the eigenstate varies periodically with time period
\begin{equation}
t_\mathrm{P} = \frac{\hbar}{|F|} \frac{2\pi w(\gvec{1})}{a} = \frac{w(\gvec{1})}{\beta}\frac{h}{|bF|} \label{eq:blochperiod}
\end{equation}
corresponding to the time taken for the momentum to span the Brillouin zone in the mixed direction $\gvec{1}$.

Non-trivial windings of space-time bands can also generate an overall drift of particle positions during each period of the Floquet-Bloch oscillation.
Consider a wavepacket representing a localized particle, which traverses a space-time band over the time interval $t_\mathrm{P}$.
The shift in its position  over that period is given by the integrated group velocity~\cite{gao2021spaceTimeTunneling},
\begin{equation}
  \Delta x = \int_0^{t_\mathrm{P}}\! \pdv{\omega}{k} \odif{t} = \int_k^{k+\Delta k} \! \frac{1}{|\dot{k}|}\pdv{\omega}{k}\,\odif{k}=\frac{\hbar}{|F|} \Delta \omega,
\end{equation}
where $\Delta k$ and $\Delta \omega$ are respectively the shift in quasimomentum and quasifrequency of the band when the momentum-frequency pair reaches its lattice image.
This shift can be expressed using winding numbers as
\begin{equation}
  \Delta x = \frac{\hbar}{|F|}\left[\frac{2\pi}{\tau_0} w(\gvec{0})+\frac{2\pi \rip}{\im\tau_0} w(\gvec{1}) \right] = \frac{2\pi\hbar}{|F|\tau_0} Q_{\tau_0}, \label{eq:blochosc}
\end{equation}
revealing a connection between the fractional charge pumped during the shortest time period $\tau_0$ for a completely filled band, and the drift of a localized particle comprised only of a \emph{subset} of the eigenstates of that band under a constant electric field.
A similar connection was made for Floquet band structures in Ref.~\onlinecite{Ke2020}; \eqnref{eq:blochosc} generalizes the result to space-time crystals.

We tested these predictions in simulations of Gaussian wavepackets in the tight-binding chain described in \figref{fig:sims}. 
The external force field was implemented by adding a linear on-site potential $\mathbf{H}_F = -\sum_j F j |j\rangle \langle j |$ to the Hamiltonian.
For strong fields, the Floquet-Bloch states tunnel seamlessly across the weakly avoided crossings due to Landau-Zener tunneling~\cite{Eckardt2008a}, making the winding topology implied by the diabatic band assigment in \subfigref{fig:sims}{a} accessible (see \siRefer for a detailed justification).
Accordingly, the period and drift of the simulated wavepackets match Eqs.~\eqref{eq:blochperiod} and \eqref{eq:blochosc} evaluated for the diabatic bands (\subfigref{fig:sims}{c}), except at the largest field value where the adiabatic evolution assumption breaks down.

\paragraph{Nonreciprocal wave transport.} \label{sec:nonreciprocal}

Traveling-wave modulation breaks spatial inversion symmetry, enabling nonreciprocal response~\cite{Cullen1958, nassar2020nonreciprocity} and directional amplification~\cite{zangeneh-nejad2019review, nassar2020nonreciprocity, Yin2022} in wave systems.
These phenomena are typically studied using frequency-domain approaches~\cite{Cullen1958,Hessel1961,Cassedy1963,Cassedy1967,Yu2009,Zanjani2014,Swinteck2015,Trainiti2016}, with  eigenmodes expressed using a discrete but infinite set of Fourier components that are coupled by the parametric modulation.
Practical calculations keep a limited subset of components (similar to techniques for quantum systems~\cite{Rudner2020a,gao2021spaceTimeTunneling,peng2022spaceTimeCrystal}), but such perturbative approaches can break down under strong drive or interactions.
Generalizing our previous results on directional amplification in a ring geometry~\cite{melkani2024resonance}, we show that the space-time band structure provides a formally exact description of nonreciprocal wave transport based on time-evolution over $\tau_0$ that, unlike its Floquet counterpart, explicitly anticipates nonreciprocity and avoids spurious band crossings.

\begin{figure}[tb]
\includegraphics{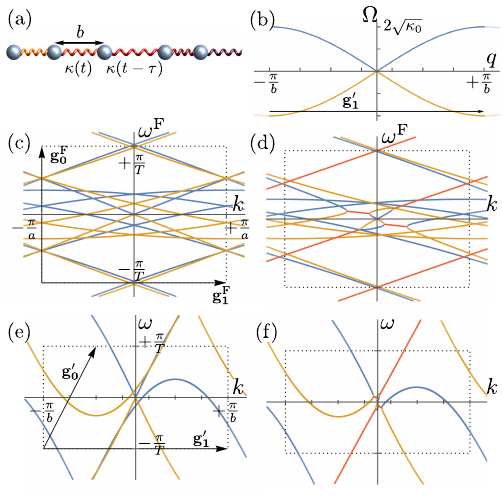}
\caption[]{\label{fig:noah}
\subfiglabel{a} Spring-mass chain with coupling springs modulated according to \eqnref{eq:modulation} (here $\alpha =2, \beta =7$).
\subfiglabel{b} Static band structure is invariant on translation by reciprocal vector $\gvec{1}' = (\frac{2\pi}{b},0)$.
\subfiglabel{c} Floquet band structure in $\delta \to 0$ limit, obtained by folding \subfiglabel{b} along   $\gvecF{0} = (0,\frac{2\pi}{T})$,  $\gvecF{1} = (\frac{2\pi}{7b},0)$.
\subfiglabel{d}  Floquet bands for $\delta = 0.5$. Amplified modes ($\operatorname{Im}(\omega) \neq 0$) in red.
\subfiglabel{e, f} Space-time band structure at infinitesimal and finite modulation.
}
\end{figure}

We illustrate these features in a 1D spring-mass chain with a traveling-wave modulation that generates broadband parametric resonance in its vibrational spectrum~\cite{kruss2022oneway}.
The chain is formed by connecting unit masses with springs of rest length $b$ and base stiffness $\kappa_0$. 
Upon modulating the $j$th spring stiffness as 
\begin{equation} \label{eq:modulation}
\kappa_j(t) = \kappa_0 \left[1 + \delta \cos \left ( \frac{2\pi \ip}{\im b}j - \frac{2\pi}{T} t\right) \right],
\end{equation}
a stiffness wave with speed $v_\textrm{m} = \frac{\im b}{\ip T} = \frac{b}{\ip \tau_0}$ is set up in the chain  (\subfigref{fig:noah}{a}) that singles out a particular direction and thus breaks reciprocity.
The ensuing equations of motion can be cast into the form of a time-dependent Schr\"odinger equation with a non-Hermitian space-time symmetric Hamiltonian~\cite{kruss2022oneway, melkani2024resonance}.

The static chain ($\delta \to 0 $) has a dispersion relation which is linear as $q \to 0$, $\Omega_{\pm}(q) \approx \pm v_\textrm{s} q$ where $v_\textrm{s} := \sqrt{\kappa_0}$ is the speed of sound in the chain.
To anticipate the effect of modulation, we fold the static band structure along the reciprocal lattice vectors of the effective Hamiltonian; coincidences in folded eigenvalues signify resonances that generate parametrically amplified modes when the modulation is turned on~\cite{YakubovichStarzhinskii,melkani2024resonance}. 
In the Floquet framework (as in Ref.~\onlinecite{kruss2022oneway}), the folding is dictated by the rectangular reciprocal lattice vectors $\gvecF{0}$ and $\gvecF{1}$ (\subfigref{fig:noah}{c}).
However,  since these vectors are symmetric with respect to $k \mapsto -k$ symmetry, the folding also respects this symmetry.
Consequently, the nonreciprocal response of the modulated spectrum is not anticipated in the Floquet approach (notice the $k \mapsto -k$ inversion symmetry in \subfigref{fig:noah}{c} which is actually broken upon evaluating the Floquet spectrum \subfigref{fig:noah}{d}).
The Floquet band structure exhibits extraneous band crossings that do not translate into resonances.

In contrast, the reciprocal lattice vectors of the space-time band structure lie along mixed frequency and momentum directions, and explicitly incorporate the breaking of inversion symmetry.
For convenience, we choose reciprocal lattice primitive vectors $\gvec{0}' = \left(\frac{2\pi\alpha}{\beta b}, \frac{2\pi}{\beta \tau_0} \right) = \alpha \gvec{1} + \rim \gvec{0}$ and  $\gvec{1}' = \left(\frac{2\pi}{b}, 0 \right) = \im \gvec{1} -\rip \gvec{0}$.
Folding the static band structure along these vectors  (\subfigref{fig:noah}{e})  gives an inversion-asymmetric band structure that anticipates amplification only for rightward-moving waves (dispersion with positive slope)~\cite{kruss2022oneway,melkani2024resonance} while avoiding any extraneous crossings.
The overlap in the linear part of the spectrum is maximized when the slope of $\gvec{0}'$, i.e., the modulation wave speed $v_\textrm{m}$, equals the sound speed $v_\text{s}$ (the so-called sonic limit~\cite{Hessel1961,Galiffi2019,kruss2022oneway}).
The anticipated resonance is confirmed by the exact evaluation of the space-time band structure for $\delta > 0$ (\subfigref{fig:noah}{f}): nearly the entire rightward-propagating band is amplified.
Frequency-domain approaches can become inaccurate near such broadband instabilities~\cite{Hessel1961}; in contrast, the space-time bands preserve the accuracy of the direct Floquet approach while correctly anticipating the nonreciprocal response.

\paragraph{Conclusion.}
The space-time Floquet operator we have introduced enables analysis of space-time crystals through direct time-evolution, avoiding truncation ambiguities inherent to frequency domain methods.
This approach applies across platforms including natural~\cite{wang2013observation} and artificial~\cite{Lohse2015,Nakajima2015} quantum systems, photonic crystals~\cite{Rechtsman2013} and active metamaterials~\cite{Yin2022}.
The two windings of its oblique bands provide a unified description of fractional charge transport and the period and drift of Floquet-Bloch oscillations, while the band representation exposes nonreciprocal amplification without frequency-domain truncation ambiguities.
The relationship between windings and transport phenomena, which is exact for isolated bands,  persists with small quantifiable corrections for band structures with weakly avoided crossings.
Promising future directions include extending our analysis to systems with synthetic dimensions~\cite{Martin2017a,Upreti2020,Adiyatullin2023}, connecting Berry phases of space-time band eigenfunctions to physical observables~\cite{Lieu2018}, analyzing the effect of gauge fields~\cite{zhang2023projective}, and investigating the non-Hermitian skin effect and non-Bloch theory~\cite{Yokomizo2019,Zhang2022c,Matsushima2025} for the space-time band structure.
The algebraic relation \eqnref{eq:Floquetfactorize} suggests connections to $n$-th root topological insulators~\cite{arkinstall2017squareRoot,marques2021nRoot} and sublattice~\cite{Asboth2013,Asboth2014,Fruchart2016} and time-glide symmetry~\cite{Morimoto2017,Chaudhary2020}.

\paragraph{Data Availability.}
All graphs were generated by numerical evaluation of the mathematical operations described in the text and can be reproduced using standard scientific computing software. 
The specific code used to generate the figures in the manuscript is available from the authors upon reasonable request.

\nocite{melkani2024resonance, kitagawa2010periodic, dana1985hall, thouless1982hall, avron2014diophantine, xu2018spaceTimeGroup, Hone1997, Eckardt2008a, Eckardt2015,Shih1994, Avron1999, Privitera2018, Citro2023, Lindner2017, gao2021spaceTimeTunneling, Eckardt2008, Breuer1989, Hijii2010, steck2025quantum, Jurgensen2021}

\paragraph{Acknowledgments.}
  \acktext

\setcounter{figure}{0}
\renewcommand{\thefigure}{A\arabic{figure}}

\section{Derivation of the space-time Floquet theorem}\label{app:spacetime}

The time-propagation operator generated by the Hamiltonian evolution satisfies
\begin{align}\label{eq:spaceTimeDerivation}
	U(t+ \tau_0) &= \mathcal{T}\bracks*{ e^{ -\iunit \int_{\tau_0}^{t+\tau_0} \odif{t'} \, H(t')} }U(\tau_0),\nonumber \\
	&= \mathcal{T}\bracks*{ e^{ -\iunit \int_{0}^{t} \odif{t'} S^{-1} \parens*{\rip b} H(t') S \parens*{\rip b} }} U(\tau_0),\nonumber \\
	&= S^{-1} \parens*{\rip b} U(t) S \parens*{\rip b} U(\tau_0),
\end{align}
where $\mathcal T$ denotes the time-ordering operator.
The last relation, which is \eqnref{eq:spaceTimeFloquet} of the main text, generalizes Floquet's theorem to space-time systems.
If $S(\rip b)$ were identity such that $\ham(t) = S(\rip b)\ham(t+\tau_0)S^{-1}(\rip b) = \ham(t+\tau_0)$, then this derivation would recover Floquet's theorem.

\section{Defining the space-time Floquet operator} \label{app:stfo}

We first change basis to the eigenstates of $S(a)$ which simultaneously block-diagonalizes the operators $\ham(t)$, $U(t)$, and $S(\rip b)$ into $N$-dimensional blocks where $N$ is the number of system degrees of freedom in a supercell of size $a$.
For this, choose a basis of states that simultaneously block-diagonalizes the matrices $\ham(t)$, $U(t)$, and $S(\rip b)$.
The eigenstates of $S(b)$ which are Bloch states with $\Psi(x+b) \propto \Psi(x)$ provide such a basis.
In particular, they are also the eigenstates of $S(a) = S^{\im}(b)$ and satisfy $\Psi(x+a) = e^{\iunit k a} \Psi(x)$ with $k$ being the Bloch wave vector.
The Bloch wave vectors $k$ take $n$ unique values given by $ \frac{2\pi j}{n a}$ with $j \in \{0,1,\dots, n-1\} = \mathbb{Z}_n$.
In the continuum limit ($n\to\infty$), we shall take the wave vector $k$ to be a real number $\in (-\frac{\pi}{a}, \frac{\pi}{a} ]$.

Since both $\mathcal{H}(t)$ and $U(t)$ commute with $S(a)$, they can be block-diagonalized in the basis of these eigenstates.
The block matrices so formed act on the uncoupled Bloch subspaces.
Thus,
\begin{equation}
\mathcal{H}(t) = \mathcal{H}_0(t) \oplus \dots \oplus \mathcal{H}_{n-1}(t) = \mathop{\oplus}\limits_{k} \mathcal{H}_k(t),
\end{equation}
$U(t) = \mathop{\oplus}\limits_{k} U_k(t)$, and $S\parens*{\rip b} = \mathop{\oplus}\limits_{k} S_k\parens*{\rip b}$. 
Here, the operator $\oplus$ denotes the direct sum and matrices labelled by the wave vector $k$ denote the different block matrices.
Using $\parens*{\mathop{\oplus}\limits_{j} X_j}\parens*{\mathop{\oplus}\limits_{j} Y_j} = \mathop{\oplus}\limits_{j} \parens*{X_j Y_j}$ on \eqnref{eq:spaceTimeFloquet} we get
\begin{align}
U_k\parens*{t+ \tau_0} &= S^{-1}_k\parens*{\rip b} U_k(t) S_k\parens*{\rip b} U_k\parens*{\tau_0}
\end{align}
and applying this $\im$ times,
\begin{align}
U_k(T) = U_k\parens*{\im \tau_0} &= S_k^{-1}\parens*{\rip a} \bracks*{S_k \parens*{\rip b} U_k\parens*{\tau_0}}^\im,\\
&= e^{- \iunit \rip k a} \bracks*{S_k\parens*{\rip b} U_k\parens*{\tau_0} }^\im.
\end{align}
The last equation results from $S_k(a) = e^{\iunit k a} \mathbb{I}$ in the chosen basis.
Taking the principal $\im$th root of $e^{- \iunit \rip k a}$ we define
\begin{equation}\label{eq:defineSTblock}
X_k := e^{- \iunit \rip k b}S_k(\rip b) U_k(\tau_0)
\end{equation}
as the $k\textrm{th}$ block of the \emph{space-time Floquet operator}, i.e., the one acting on the Bloch subspace corresponding to wave vector $k$.
It is invariant on translation by the  primitive vectors $\vb a_0$ and $\vb a_1$, thus capturing the symmetries of the system.
The important result is that each Bloch block of the Floquet operator can be factorized as 
\begin{equation}
	U_k(T) = X_k^\im.\label{eq:FloquetfactorizeApp}
\end{equation}

Note that for the case of a space-time symmetric ring~\cite{melkani2024resonance}, we have $(n=1)$ and \eqnref{eq:translationalSymmetry} only expresses the periodic boundary conditions of the system.
In other words, the translation $S(a)$ is over the whole system length and due to PBC is identity.
In this case, it is not possible to block-diagonalize the operators and we use
\begin{align}
U(T) = U\parens*{\im \tau_0} &= S^{-1}\parens*{\rip a} \bracks*{S\parens*{\rip b} U\parens*{\tau_0}}^\im,\\
&= \bracks*{S\parens*{\rip b} U\parens*{\tau_0} }^\im,
\end{align}
to get the space-time Floquet operator as $X = S(\rip b) U(\tau_0)$---a restricted case of which was derived in Ref.~\cite{melkani2024resonance}.
We cannot have a band structure in this case of a ring geometry.

\section{Derivation of quantized current}\label{app:quantized}

For a Hermitian lattice system, the total pumped current integrated up to some time period $t'$ is~\cite{kitagawa2010periodic}
\begin{equation}\label{eq:integratedCurrent}
	\int_0^{t'} \odif t \, J(t) = \int_{\mathrm{BZ}} \frac{\odif k}{2\pi} \iunit \Tr \bracks*{P_k U_k^\dagger(t') \pdv{U_k}!{k}(t')}.
\end{equation}
Here $P_k$ is the projector onto the occupied band or the bands whose transport is being evaluated.
In the derivation below we assume unitary dynamics so that $U_k^\dagger=U_k^{-1}$.

Quantization of current for a Hermitian Floquet system can be derived by integrating \eqnref{eq:integratedCurrent} up to the time period $T$.
To see this, let $\ket{u_{k,j}}$ denote a Floquet state with Bloch wave vector $k$ from the occupied Floquet band with respect to which we are taking the trace.
It satisfies $U_k(T)\ket{u_{k,j}} = e^{-\iunit \omega_j^{\textrm{F}}(k) T} \ket{u_{k,j}}$.
Differentiating this (and suppressing the quantum numbers of the states) we get
\begin{equation}
	\pdv{U_k(T)}!{k}\ket{u} = e^{-\iunit \omega^{\textrm{F}} T} \parens*{\pdv{\ket{u}}!{k} -\iunit T \pdv{\omega^{\textrm F}}!{k}\ket{u}} - U_k(T)\pdv{\ket{u}}!{k}\nonumber
\end{equation}
and thus
\begin{equation}
	\bra{u}U^{-1}_k(T)\pdv{U_k(T)}!{k}\ket{u} = -\iunit T \pdv{\omega^{\textrm F}}!{k}.
\end{equation}
Substituting this in \eqnref{eq:integratedCurrent} we have
\begin{align}
	\int_0^{T} \odif t \, J(t) &= \int_{\mathrm{BZ}} \frac{\odif k}{2\pi} \, \iunit \Tr \bracks*{P_k U^\dagger_k(T) \pdv{U_k}!{k}(T)},\\ 
	&= T \int_{\mathrm{BZ}} \frac{\odif k}{2\pi} \, \sum_j \pdv{\omega_j^{\textrm F}(k)}!{k},\\
	&= w(\vb g_0^{\textrm F}).
\end{align}
In the second to last line we sum over all Floquet states with Bloch wave vector $k$ and unique state indices $j$ that belong to the same band.

If multiple bands are occupied, the result is the sum of the winding numbers of these bands.
This conventional calculation uses the full period because $U_k(T)$ is the operator that closes the temporal cycle of a generic driven lattice.
It therefore produces an integer winding on the rectangular Floquet Brillouin zone and contains no distinguished invariant associated with an arbitrary subinterval $T/\beta$.

Now, for a space-time symmetric system we have from the definition of the space-time Floquet matrix, \eqnref{eq:defineSTcont},  
\begin{align*}
	X_k^{-1}\pdv{X_k}!{k} &= U_k^{-1}(\tau_0)\pdv{U_k(\tau_0)}!{k} - \iunit \rip b\\ 
	&+U_k^{-1}(\tau_0)S_k^{-1}(\rip b)\pdv{S_k(\rip b)}!{k} U_k(\tau_0).
\end{align*}
This time, let $\ket{u_{k,j}}$ denote the space-time eigenvector with Bloch wave vector $k$ from the occupied space-time band with respect to which we are taking the trace.
Using the stroboscopic dynamics of \eqnref{eq:stroboscopic} we have (suppressing the quantum numbers in the notation again)
\begin{align*}
\bra{u} U_k^{-1}(\tau_0)S_k^{-1}(\rip b)&\pdv{S_k(\rip b)}!{k} U_k(\tau_0)\ket{u} = \\ &\bra{u}\pdv{S_k(\rip b)}!{k} S_k^{-1}(\rip b)\ket{u}.
\end{align*}

To evaluate this quantity, we first note that any wave vector in the $k$th Bloch subspace can be written as $q_n = k+nK$, with $n \in \mathbb{Z}$ and $K = 2\pi/a$.
The translation operator acts on the plane waves diagonally as $S(r)\ket{q_n}=e^{\iunit q_n r}\ket{q_n}$.
Using the fixed cell-periodic Bloch basis, the projected translation matrix on the $k$th Bloch subspace can be written as
\begin{equation}
S_k(r)=e^{\iunit kr}D(r),
\end{equation}
where
\begin{equation}
D(r)_{nn'}=e^{\iunit nKr}\delta_{nn'}
\end{equation}
is independent of $k$ such that $\pdv{S_k(r)}!{k}S_k^{-1}(r)=\iunit r\,\mathbb{I}$ for any $r$.
Thus, for any normalized state $\ket{u}$ in the $k$th Bloch subspace,
\begin{equation}
\bra{u}\pdv{S_k(\rip b)}!{k} S_k^{-1}(\rip b)\ket{u} = \iunit r_\alpha b
\end{equation}

Substituting this,
\begin{align}
	\int_0^{\tau_0} \odif t \, J(t) &= \int_{\mathrm{BZ}} \, \frac{\odif k}{2\pi} \iunit \Tr \bracks*{P_k U_k^\dagger(\tau_0) \pdv{U_k}!{k}(\tau_0)},\\ 
	&= \int_{\mathrm{BZ}} \, \frac{\odif k}{2\pi} \iunit \Tr \bracks*{P_k X_k^{-1} \pdv{X_k}!{k}},\\
	&= \tau_0 \int_{\mathrm{BZ}} \frac{\odif k}{2\pi} \, \sum_j \pdv{\omega_j (k)}!{k}, \label{eq:si-intgroupvel}\\
	&= w\parens*{\vb g_0} + \frac{\rip}{\im} w\parens*{\vb g_1}. \label{eq:si-pumpcurrent}
\end{align}

Thus the current integrated up to a duration $\tau_0$ is quantized to a fractional value determined by the windings of the space-time band structure.
The fractional term originates from evaluating the winding on the oblique space-time reciprocal lattice, where translation along $\gvec{1}$ changes both wavevector and frequency.
This is the sense in which the result does not follow directly from the ordinary Floquet winding over a full temporal period.

As a consistency check, the current integrated up to the full Floquet period $T = \im \tau_0$ is $\beta$ times the above value, i.e., $\beta w\parens*{\vb g_0} + \rip w\parens*{\vb g_1}$.
This should equal the result derived via Floquet theory, namely $w(\vb g_0^{\textrm F})$.
To see this we first note that $\gvec{0} = \im \gvecF{0}$ and $\gvec{1} = \rip \gvecF{0} + \gvecF{1}$. 
Since the Floquet operator is obtained by applying the space-time Floquet operator $\beta$ times in succession, the corresponding Floquet band then satisfies
\begin{align}
(k, \omega^{\textrm F}) &= (k, \omega^{\textrm F}) + w(\gvec{0}) \beta \gvecF{0} + w(\gvec{1}) \parens*{\rip \gvecF{0} + \gvecF{1}}, \nonumber \\
&= (k, \omega^{\textrm F}) + \left [\beta w(\gvec{0}) + \rip w(\gvec{1}) \right ] \gvecF{0} + w(\gvec{1}) \gvecF{1},\nonumber
\end{align}
and we identify $w(\vb g_0^{\textrm F}) = \beta w(\gvec{0}) + \rip w(\gvec{1})$ as the vertical and $w(\vb g_1^{\textrm F}) = w(\gvec{1})$ as the horizontal winding number for the Floquet band.
This Diophantine-like constraint on the vertical Floquet winding number in a space-time symmetric system is reminiscent of a similar constraint on the form of the Chern number governing the quantum Hall conductance~\cite{dana1985hall, thouless1982hall, avron2014diophantine}.

\section{The space-time Floquet operator in higher dimensions} \label{sec:higherd}

The framework of defining a space-time Floquet operator can be generalized to $(D+1)$ dimensions.
Consider a system with periodic boundary conditions in all $D$ spatial directions, governed by a Hamiltonian of the form $\mathcal{H}(t) = \frac{\vb p \cdot \vb p}{2m} + V(\vb x,t)$.
As we showed explicitly for $D=1$ above, imposing periodic boundary conditions  force $V(\vb x,t+T) =V(\vb x, t)$ for some time period $T$.
In the simplest case (in the absence of symmetries involving point-group operations or time-reversal~\cite{xu2018spaceTimeGroup}), a space-time crystal's $D+1$ primitive lattice vectors are given by
\begin{subequations}
\begin{align}
\vb{\Tilde{a}_i} &= \left( \vb b_i, \frac{\alpha_i}{\beta_i}T \right) \quad \text{ for } \quad i = 1,\dots, D, \\
\vb{\Tilde{a}_0} &= \left( 0, T \right).
\end{align}
\end{subequations}
Here, the $\vb b_i$ are linearly independent $D$ dimensional vectors and $\alpha_i,\beta_i$ are pair-wise co-prime.
Concretely, the potential satisfies $V(\vb x + \vb b_i,t + \frac{\alpha_i}{\beta_i}T) = V(\vb x,t)$ and is periodic in time $V(\vb x ,t + T) = V(\vb x,t)$.

Our strategy is again to identify the smallest time-step governing the dynamics, and use it to define new primitive lattice vectors $\vb a_i$ that satisfy the space-time symmetry. 
We denote these by $\vb a_i$ for $i=0,\dots, D$.

For $i=1,\dots, D$, we can form $\vb a_i = \beta_i \vb{\Tilde{a}_i} - \alpha_i \vb{\Tilde{a}_0} = \left( \beta_i \vb b_i, 0 \right)$.

To form the last one, we note that for a given pair of coprime integers $\alpha_i$ and $\beta_i$, integers $s_i$ and $q_i$ exist such that $s_i\alpha_i + q_i \beta_i = 1$ (by B\'{e}zout's identity).
Thus, we have
\begin{equation}
s_i\vb{\Tilde a_i} + q_i \Tilde{\vb a_0} = \left( s_i \vb b_i, \frac{T}{\beta_i} \right).
\end{equation}

Now, by the generalized B\'{e}zout's identity, integers $p_i$ exist such that 
\begin{equation}
\sum_{i=1}^D p_i \frac{1}{\beta_i} = \frac{1}{\operatorname{lcm}(\beta_1,\dots,\beta_D)} =: \frac{1}{l}.
\end{equation}
The vector $\vb a_0$ is then given by
\begin{equation}
\vb a_0 = \sum_{i=1}^D p_i \left( s_i \vb{\Tilde a_i} + q_i \vb{\Tilde a_0} \right) = \left(\sum_{i=1}^D p_is_i \vb b_i, \frac{T}{l} \right).
\end{equation}

That is, the smallest time-step is $\tau_0 := T/l$ where $l$ is the least common multiple of the integers $\beta_i$.
The resulting primitive lattice vectors for the space-time lattice are
\begin{subequations}
\begin{align}
\vb a_i &= \left( \beta_i \vb b_i, 0 \right) \quad \text{ for } \quad i = 1,\dots, D, \\
\vb a_0 &= \left(  \vb b_0 , \tau_0 \right) := \left(\sum_{i=1}^D p_is_i \vb b_i, \tau_0 \right).
\end{align}
\end{subequations}

The system enjoys the following symmetries:
\begin{subequations}
	\begin{align}
	\ham(t) &= S(\beta_i \vb b_i)\ham(t)S^{-1}(\beta_i \vb b_i) \quad \text{ for } i=1,\dots, D,\\
	\ham(t) &= \ham(t+T), \\
	\ham(t) &= S(\vb b_0)\ham(t+\tau_0)S^{-1}(\vb b_0), 
	\end{align}
\end{subequations}
corresponding to the $D$ translational symmetries, the Floquet symmetry, and the space-time symmetry respectively.
As in the $D=1$ case, any of the $\beta_i \vb b_i$ can be of the extent of the full system in the respective direction, in which case the corresponding equation expresses the periodicity in the boundary conditions rather than a translational symmetry.

With these lattice vectors defined, we now choose a basis of Bloch vectors which are simultaneous eigenstates of the translation operators $S(\beta_i \vb b_i)$ and thus satisfy
\begin{equation*}
  S(\beta_i \vb b_i)\Psi_{\vb k}(x) = e^{\iunit \beta_i \vb b_i \cdot \vb{k}}\Psi_{\vb k}(x)
\end{equation*}
with $\vb k$ denoting the $D$-dimensional wavevector indexing the distinct Bloch states. 
This basis block-diagonalizes all of our matrices of interest providing us the block matrices $U_{\vb k}(t)$ and $S_{\vb k}(\vb b_i)$.
Following the same algebraic steps as in the $D=1$ case,  we find that the time-evolution operator over the smallest time interval has the form
\begin{equation}
	U_{\vb k}(t + \tau_0) = S_{\vb k}^{-1}(\vb b_0) U_{\vb k}(t) S_{\vb k}(\vb b_0) U_{\vb k}(\tau_0).
\end{equation} 
Upon applying this evolution $l$ times, we obtain the Floquet operator
\begin{equation} \label{eq:uTdplusone}
	U_{\vb k}(T) = S_{\vb k}^{-l}(\vb b_0) \bracks*{S_{\vb k}(\vb b_0) U_{\vb k}(\tau_0)}^l.
\end{equation} 
We show in the bottom of this section that the translation along the primitive lattice direction $\vb b_0$ can be simplified as  $S_{\vb k}^{-l}(\vb b_0) = e^{-\iunit l \vb b_0 \cdot \vb k}$.
Using this result in \eqnref{eq:uTdplusone} and taking the principal $l$th root, we obtain $U_{\vb k}(T) = X_{\vb k}^l$ where
\begin{equation}
	X_{\vb k} = e^{-\iunit \vb b_0 \cdot \vb k} S_{\vb k}(\vb b_0) U_{\vb k}(\tau_0)
\end{equation}
is the space-time Floquet operator for the system.

We leave generalization to other crystal groups such as those involving point-group operations or time-reversal~\cite{xu2018spaceTimeGroup} for future work.

To complete the derivation, we now show that $S_{\vb k}^{-l}(\vb b_0) = e^{-\iunit l \vb b_0 \cdot \vb k}$.
Since $l$ is the lowest common multiple of the $\beta_i$, there exist integers $\gamma_i:= l/\beta_i$ which we now use to simplify the expression for $S_{\vb k}^{-l}(\vb b_0)$.
\begin{align}
	S_{\vb k}^{-l}(\vb b_0) &= S_{\vb k}^{-1} \parens*{l \sum_{i=1}^D p_i s_i \vb b_i},\\
	&= S_{\vb k}^{-1}\parens*{\sum_{i=1}^D \gamma_i p_i s_i \beta_i \vb b_i}, \\
	&= \prod_{i=1}^D S_{\vb k}^{-1}(\gamma_i p_i s_i \beta_i \vb b_i), \\
	&= \prod_{i=1}^D \bracks*{e^{\iunit \beta_i \vb b_i \cdot \vb k}}^{-\gamma_i p_i s_i}, \\
	&= e^{-\iunit l \sum_{i=1}^D p_i s_i \vb b_i \cdot \vb k},\\
  &= e^{-\iunit l \vb b_0 \cdot \vb k}.
\end{align}
We used $S_{\vb k}^{-1}(\beta_i \vb b_i) = e^{-\iunit \beta_i \vb b_i \cdot \vb k} \mathbb{I}$ in the third to last line.

\section{Weakly avoided crossings and robustness of quantized transport}

As \subfigref{fig:fig1}{c,d} shows, space-time bands with nontrivial windings along both reciprocal directions may intersect other bands or themselves. This section qualifies the numerical demonstrations of the fractional transport result: the winding formula is exact for isolated space-time bands, while the tight-binding examples below contain extremely small avoided crossings that produce controlled, quantifiable deviations.
According to the Wigner-von Neumann theorem, in the absence of fine-tuning such intersections will generically lead to avoided crossings with vertical gaps in the frequency direction.
However, when the coincidences do not correspond to direct resonances between pairs of bands separated in frequency-wavevector space by the modulation wavevector, they are expected to be small.
The existence of these gaps in (1+1)D Floquet spectra, termed weakly avoided crossings (WACs)~\cite{Hone1997}, and their relevance (or lack thereof) for physical measurables has been appreciated previously in the context of periodically-driven cold atom systems~\cite{Eckardt2008a,Eckardt2015} and Thouless pumping~\cite{Shih1994,Avron1999,Privitera2018,Citro2023}.

In a perturbative description, WACs reflect higher-order couplings between eigenstates of the static system due to the time-dependent part of the Hamiltonian, in contrast to the primary gaps that arise from first-order couplings~\cite{Lindner2017}.
WACs impact the dynamics at frequency scales set by their associated gaps, which are far slower than the characteristic frequency scales of the modulation $\sim 1/\tau_0$ or the primary gaps $\sim \hbar/V_0$.
As a result, signatures of nontrivial windings can in practice be seen even in the presence of WACs in both Thouless pumping~\cite{Shih1994,Privitera2018,Citro2023} and Floquet-Bloch oscillations~\cite{gao2021spaceTimeTunneling}, with quantifiable corrections for both transport mechanisms due to the WACs that restrict the parameter ranges within which the winding predictions are accurate.
In our work, we use a tight-binding model for numerical tests of our results which leads to band crossings and WACs.
We follow the approach of these prior works to argue that our topological predictions, which are exact in the case of bands without any crossings, become approximate when WACs are taken into account, and the corrections are quantifiably small.

\subsection{Band assignment}

\begin{figure*}
  \centering
  \includegraphics{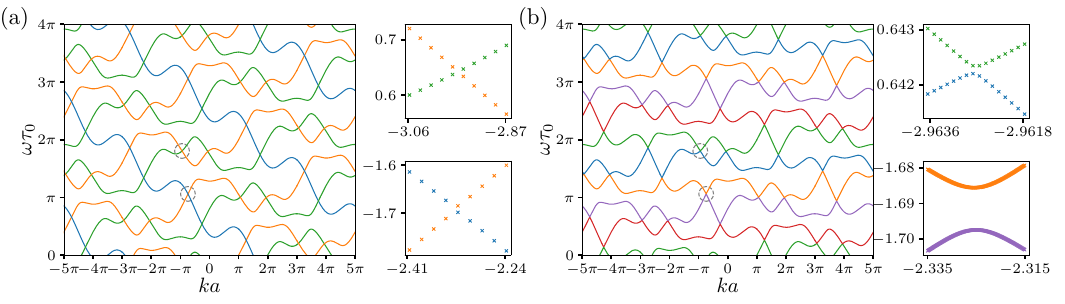}
  \caption{Numerical band assignment is sensitive to sampling resolution of wavevector space. We show numerical space-time band structures for the tight-binding model from the main text, evaluated with different discrete wavevector values along the BZ.
    The inversion symmetry $V(j,t) = V(-j,-t)$ places the space-time band structure in wallpaper group $p2$~\cite{xu2018spaceTimeGroup}, so there are only two unique apparent crossing points (dashed circles).
    \subfiglabel{a} Space-time band structure generated by the eigenvector continuity criterion with the BZ spanned by $n=300$ wavevector values (spacing  $\Delta k = \pi/(300a)$ between successive wavevectors). Right panels are zooms in the vicinity of the two circled crossings, showing the discrete values.
    \subfiglabel{b} Same as (a) for $n=50,000$ wavevector values. When WACs are taken into account, the five subspaces unite to a single space-time band; colors indicate different Floquet-Bloch replicas of the contiguous band. 
  }
  \label{fig:bandassignment}
\end{figure*}

In our numerical evaluation of the space-time band structures, we evaluate the spectrum of the space-time Floquet operator at discrete values of the wavevector $k$, which provides $N$ eigenvalues and their corresponding wavevectors at each $k$.
To assign band indices to these eigenstates, we may use two approaches: eigenvalue or eigenvector continuity.
For eigenvalues, we start with an arbitrary assignment at the first $k$ value, identify the nearest eigenvalue and then use the notion of eigenvalue continuity: at each successive $k$ value, eigenstates are assigned state indices that minimizes the mean squared distance between the true eigenvalue and a prediction of the expected eigenvalue for that band based on a linear extrapolation from the previous two.
For eigenvectors, we assign bands based on the maximal overlap of the eigenvectors at each $k$ value with the eigenvectors at the preceding $k$ value.
We verified that both approaches give the same results for the bands shown in the text.

We found that the band assignment based on continuity depends on the spacing of the sampled wavevector points.
If the spacing between successive wavevectors is large enough so that the gap between successive eigenvalues is larger than any WACs, the bands appear to cross without any gaps opening up, as shown for space-time bands evaluated with $n=300$ points spanning the BZ in \subfigref{fig:bandassignment}{a}.
This is consistent with the influence of the WAC being highly localized in both frequency and wavevector space.
However, if wavevector space is sampled finely enough, then the continuity assignment follows the expectation of an avoided crossing.
For the tight-binding model from the main text, the larger WAC becomes apparent when evaluating the space-time band structure with $n=500$ values, but the smaller WAC can only be resolved if the resolution of the numerical evaluation is increased by a factor of over 100 compared to \subfigref{fig:bandassignment}{a} (note much smaller $ka$ and $\omega\tau_0$ ranges in \subfigref{fig:bandassignment}{b}, for which $n=50,000$ discrete wavevector values were used).
When all WACs are correctly sampled to approximate the true band structure in the $n \to \infty$ limit, the five subspaces of the Bloch eigenstates unite to form a single band with no net winding in the frequency direction.
These avoided crossings prevent the accumulation of windings around the frequency direction.

To probe the physics of nontrivial windings in a simple system, we used the coarse sampling of wavevector space to assign bands, calculate windings, and prepare initial conditions for the tight-binding simulations; this is equivalent to ignoring the WACs when making band assignments.
We now evaluate the effect of ignoring WACs on the quantized transport simulations.

\subsection{Corrections to Thouless pumping}
The ``diabatic'' band assignment which ignored the WACs (\subfigref{fig:bandassignment}{a} and main text) generates three bands with windings $(w\parens*{\gvec{0}},w\parens*{\gvec{1}}) = (-1,1)$, $(-1,2)$ and $(-1,2)$ for the blue, orange, and green bands respectively.
Each of these diabatic bands is a collection of segments that are strictly disconnected under the true (adiabatic) band assignment shown in \subfigref{fig:bandassignment}{b}. 
As a result, the quantized current expression \eqnref{eq:si-pumpcurrent}, which is exact for isolated connected bands, becomes an approximation for the diabatic bands.
The deviation between the diabatic pumped charge and the winding-number prediction is set by the size of the WACs~\cite{Shih1994,Privitera2018}, which are exponentially suppressed as the adiabatic limit is approached~\cite{Lindner2017}.
Therefore, we expect to observe near-perfect fractional charge quantization for the diabatic bands when the modulation period is long compared to the time scales associated with the primary band gaps and band widths of the unmodulated system, with small quantifiable deviations due to the WACs that vanish in the $T \to \infty$ limit.

To see why the corrections to exact quantization are small, consider the blue diabatic band, for which the adiabatic expression \eqnref{eq:si-pumpcurrent} predicts a pumped charge $Q_{\tau_0} = -1 + 3/5 = -2/5$.
This prediction arose from integrating the group velocity of a space-time band [\eqnref{eq:si-intgroupvel}], which evaluates to a rational fraction for a contiguous band.
However, ``filling'' the diabatic blue band corresponds to populating the Floquet-Bloch eigenstates over two smooth segments separated by two WACs at the points of intersection with the green and orange bands.
The diabatic pumped charge over one cycle is still obtained by integrating the group velocity, but the integral now must be carried out piecewise over these two segments. 
The adiabatic winding expression therefore overestimates the true pumped charge by including contributions from the WAC gaps that are absent in the diabatic band.
The contribution of the $j$th gap of  width $\Delta \omega_j$ (signed according to the direction of the step in quasifrequency as $k$ advances) to the integral in \eqnref{eq:si-intgroupvel} is simply $\tau_0 \Delta \omega_j$.
The actual pumped charge of the diabatic band is therefore 
\begin{equation}
	Q_{\tau_0} - \sum_j \tau_0 \Delta \omega_j, \label{eq:si-diabaticpump}
\end{equation}
where the sum is over all gaps.
The blue band has two such gaps of size $\lesssim 0.01/\tau_0$ and negative sign, so we expect a fractional reduction in the magnitude of the pumped charge from the winding-number prediction $-2/5$ by a few percent. 
These are consistent with the measured deviations in our simulations (inset to \subfigref{fig:sims}{b} of the main text).
In the adiabatic limit, the WACs vanish and the winding expression for the pumped charge becomes exact.

We note that the band assignment made using finely-spaced points in $k$ (\subfigref{fig:bandassignment}{b}) is also consistent with the pumped-charge formula. 
The exact band structure for the tight-binding model involves a single space-time band that smoothly connects all five Bloch subspaces. 
This band has winding numbers $(w\parens*{\gvec{0}},w\parens*{\gvec{1}}) = (-3,5)$ and the quantized charged according to \eqnref{eq:si-pumpcurrent} is $-3 + 3/5 \times 5 = 0$. 
Filling the true space-time band corresponds to occupying all segments from the blue, orange, and green bands in the diabatic assignment, so the net pumped charge over a cycle is the sum of three contributions in the form of \eqnref{eq:si-diabaticpump}.
The symmetries of the space-time band structure ensure that the net signed contributions of the WACs (second term in \eqnref{eq:si-diabaticpump}) cancel, and we are left with the sum of the topological contributions $-2/5+1/5+1/5=0$.

\subsection{Floquet-Bloch oscillations from successive Landau-Zener tunnelings}

\begin{figure}
  \centering
  \includegraphics{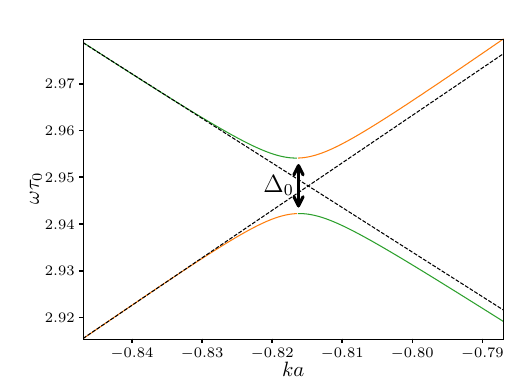}
  \caption{Largest weakly avoided crossing in Floquet-Bloch simulation.
    A zoom of the band structure in \subfigref{fig:sims}{a} is shown near the WAC between the orange and green bands.
    The gap is $\Delta_0 = 0.01187/\tau_0$. The slopes of the dispersion near the avoided crossing are obtained by performing linear fits (dashed lines) and evaluate to $\mu_1 = 1.0148 a/\tau_0$ and $\mu_2 = -0.9527/\tau_0$.
    These values provide the probability for a wavepacket in the orange (green)  band to diabatically cross the WAC and continue in the orange (green) band via \eqnref{eq:zener}.
  }
  \label{fig:zener}
\end{figure}

WACs become insignificant for a different reason in the presence of external field, as identified by early work on Floquet cold atom systems in ~\cite{Eckardt2008a,Eckardt2008} and more recently in \cite{gao2021spaceTimeTunneling} in the context of space-time crystals. 
WACs impact the dynamics at frequency scales set by their associated gaps, which are far slower than the characteristic frequency scales of the modulation $\sim 1/\tau_0$ or the primary gaps $\sim \hbar/V_0$.
As a result, a range of system parameters can often be found for which the dynamics are slow compared to the modulation time scale (so that the stroboscopic evolution provides a useful description over many intervals of $\tau_0$), yet fast compared to the WAC gap scale (so that the effect of the WACs on the dynamics is negligible)~\cite{Eckardt2008a,Eckardt2008}.
Here, we show that the system used to simulate nontrivial Floquet-Bloch oscillations lies within this regime.

The band assignment that we use, which ignored the WACs, and can be considered a high-frequency description that is accurate at scales much faster than the WAC gap scale.
The resulting Floquet-Bloch oscillations under an external field were connected to the nontrivial windings of space-time bands by assuming that the semiclassical dynamics generated by the external force field, $\dot{k} = F/\hbar$, induces the wavepacket to adiabatically follow the bands shown in \subfigref{fig:sims}{a}.
However, truly adiabatic dynamics (corresponding to $F \to 0$, i.e. extremely slow advance of the wavevector) would render the dynamics sensitive to the slow time scale associated with the miniscule gaps at the WACs, the largest of which is shown in \figref{fig:zener}.
The wavepacket would follow the connected band at the WAC, which is different from the band assigned by ignoring the WAC gap.
However, a range of force fields can be found for which the wavepacket dynamics is fast relative to the WAC scale, but slow relative to the Floquet scale.
For such forces, the states in the wavepacket are \emph{diabatically} transferred across the WAC gap by Landau-Zener tunneling~\cite{Breuer1989,Hijii2010}.
The tunneling probability $\Gamma$ can be computed by approximating the space-time band dispersions $\omega_1(k)$ and $\omega_2(k)$ near the WAC as linear with slopes $\mu_1$ and $\mu_2$ and separated by a quasifrequency gap $\Delta_0$, for which the Zener tunneling formula~\cite{steck2025quantum} gives the rate as~\cite{gao2021spaceTimeTunneling}
\begin{equation}
  \label{eq:zener}
  \Gamma = \exp \left(- \frac{\pi \hbar \Delta_0^2}{2|F(\mu_1-\mu_2)|}\right).
\end{equation}

For small gaps and/or large driving forces so that $\hbar \Delta_0^2/|F(\mu_1-\mu_2)| \ll 1$, the tunneling rate approaches one; i.e. the spectral content of the wavepacket is transferred completely across the WAC.
For the system in \figref{fig:sims}, numerical estimates of the gap and the slopes of the largest WAC  (\figref{fig:zener}) lead to  $\Gamma \approx 0.9987$ at the lowest field strength of $F = 0.003$ (in simulation units with $\hbar = 1$).
The other WACs have even higher tunneling probabilities.
Since the tunneling probability is very close to one, the wavepacket appears to adiabatically follow the space-time bands assigned by ignoring the WACs in \subfigref{fig:sims}{a}, with the period and displacements given by their associated windings on the frequency-wavevector torus as predicted in the main text.
However, for adiabaticity to be maintained at the level of the space-time bands, the traversal of the wavevector due to the field must still be slow compared to the modulation scale; i.e. $t_\mathrm{P} \gg \tau_0$ or $2\pi w(\gvec{1})\hbar/(|F|a\tau_0) \gg 1$.
At the largest force simulated, the wavepacket trajectories in \subfigref{fig:sims}{c} have begun to deviate from the predictions due to nonadiabatic effects at the larger frequency scales.

While the parameter ranges at which our quantized transport predictions are applicable are restricted by the twin conditions of slow dynamics at the modulation scale and fast dynamics at the WAC gap scale, we expect to be able to find appreciable parameter ranges that meet this restriction.
Specifically, we require the force field to satisfy
$$ \frac{\hbar \Delta_0^2}{|F(\mu_1-\mu_2)|} \ll 1 \ll \frac{\hbar}{|F|a\tau_0}.$$
When windings are order one, the slopes of the dispersion relation scale as $\mu_i \sim (1/\tau_0)/(1/a) \sim a/\tau_0$.
The restriction then becomes
$$ \Delta_0^2 \tau_0 \ll \frac{|F|a}{\hbar} \ll \frac {1}{\tau_0}.$$
Since the criterion for the gap to be ignored is that $\Delta_0 \tau_0 \ll 1$, the above condition can indeed be satisfied for a range of forces without requiring fine-tuning for the quantized transport to be observed.

\section{Numerical simulations of quantized transport}

We simulated Hamiltonian dynamics on a noninteracting tight-binding model in \textsc{Python}, with the state of a system of $n$ supercells (of  $N=\beta$ sites each) represented by a  complex vector of length $n\beta$, denoted $\Phi = \{\phi_1,\phi_2,....,\phi_{n\beta}\}$.
We calculated the space-time band structure by diagonalizing the space-time Floquet operator $X_k$ for the allowed wavevectors under periodic boundary conditions $k_p = \frac{2 \pi}{n a}p$, $1 \leq p \leq n$.
For the system parameters depicted in \figref{fig:sims}, we have $N=5$ degrees of freedom in each supercell, and the five distinct eigenvalues $\omega_j(k)$  at each $k$ were assigned indices $j \in \{1,2,3,4,5\}$ using continuity of the eigenvalues as described above.
When continuity across the space-time Brillouin zone is considered, the five states form three bands, two of which (orange and green bands in \subfigref{fig:sims}{a}) incorporate pairs of state indices.
The corresponding normalized eigenvectors $\mathbf{u}_j(k)$ were used to generate the Bloch state at the $m$th unit cell (position $x=ma$) via $\Psi_{k,j}(ma) = \frac{1}{\sqrt{N}} e^{imka}\mathbf{u}_j(k)$, normalized so that $\langle \Psi_k | \Psi_k\rangle = 1$ (with $|\mathbf{u}| = 1$).
We numerically verified that charge transport followed the quantized predictions using semiclassical  dynamics of wavepackets assembled from the Bloch states of the relevant space-time bands, as described below.

The subsequent time evolution was obtained by numerically solving the matrix differential equation (setting $\hbar = 1$)
$$\frac{d\Phi(t)}{dt} = -i \mathbf{H}(t) \Phi(t),$$
where $\mathbf{H}$ is a $(\beta n)\times(\beta n)$ array with entries $H_{ij}$ encoding the desired time-dependent tight-binding Hamiltonian.
Numerical integration was carried out using the \textsc{Python} \texttt{odeintw} package.

\subsection{Thouless pump}

For the Thouless pumping simulations, the net charge transport from all Bloch states associated with the $j$th branch of the space-time spectrum was evaluated by initializing the system in a linear superposition of the states with equal weights: $$\Phi(0) = \sum_{p=1}^n e^{ik_p na/2}\Psi_{k_p,j},$$ where the phase factor centers the resulting localized wavefunction at the center of the tight-binding chain and the sum is over the $n$ distinct $k$ values allowed under periodic boundary conditions.
This initial condition recreates a uniform population of the Bloch states from the $j$th branch, with net occupancy  $\langle \Phi | \Phi \rangle = n$~\cite{Jurgensen2021}.

We implemented the discretized current operator for the tight-binding model,
$$j_i = \frac{2}{\hbar} \text{Im}(\phi_i^* \phi_{i-1} H_{i,i-1})$$
and calculated the average instantaneous current through the chain as
$$J(t) = \frac{1}{L} \int dx\, j(x,t) \to \frac{1}{n\beta} \sum_{i=1}^{n\beta} j_i.$$
The integral of the current over discrete time steps was numerically evaluated to obtain the charge transfer $Q_j(t)$ through the chain due to all states in the $j$th subspace.
For space-time bands that combine multiple degrees of freedom at each $k$, such as the orange and green bands in \subfigref{fig:sims}{a}, the net pumped charge associated with the band was obtained by adding up the $Q_j(t)$ contributions from the constituent eigenstate indices of the band.

\subsection{Floquet-Bloch oscillations}

For the Bloch oscillation simulations, the system was initialized with a Gaussian wavepacket also centered at position $n/2$): $$\Phi(0) = \mathcal{N} \sum_{p=1}^n e^{ik_pna/2}e^{-(k_p-\bar{k})^2/(2\sigma^2)}\Psi_{k_p,j},$$
where the mean $\bar{k}$ is the carrier wavevector, the standard deviation $\sigma$ sets the width of the wavepacket in real space, and $\mathcal{N}$ is a normalization.
The parameters used for the simulations reported in the main text are $\bar{k} = -1/a$ and $\sigma = 0.3/a$.
The wavepacket position during the evolution was calculated from the state vector via
$$\langle x \rangle =\frac{\sum_{i=1}^{N\beta}  b i \phi_i^* \phi_i }{ \sum_{i=1}^{N\beta} \phi_i^* \phi_i}.$$

\end{document}